%% file: main.tex
\documentclass[conference]{IEEEtran}
\usepackage{graphicx}
\usepackage{amsmath}
\usepackage{amssymb}
\usepackage{mathtools}
\usepackage{bm}
\usepackage{subfig}
\usepackage{wrapfig}
\usepackage{comment}

\usepackage[font=small,labelfont=bf]{caption}

\usepackage{array}

\usepackage{xcolor}

\newcommand{\Liang}[1]{{\color{blue}Liang says:~#1}}

\newcolumntype{P}[1]{>{\centering\arraybackslash}p{#1}}
\newcolumntype{M}[1]{>{\centering\arraybackslash}m{#1}}

\newcommand{\ie}{{\em i.e.}}

\def\rrr#1\\{\par
\medskip\hbox{\vbox{\parindent=2em\hsize=6.12in
\hangindent=4em\hangafter=1#1}}}

\begin{document}

\input{frontmatter-sv}

\begin{abstract}
\input{abstract-rev00}
\end{abstract}

\section{Introduction}
\label{sec:introduction}
\input{introduction-rev00}

\section{Related Work}
\label{sec:relatedwork}
\input{related-work}

\section{Quantum Data Center Network}
\label{sec:problem}
\input{problem}

\section{Quantum Leaf Switch and Spine Switch}
\label{sec:solution}
\input{solution}

\section{Experiments and Evaluation}
\label{sec:evaluation}
\input{evaluation-rev00}

\section{Conclusions}
\label{sec:future}
\input{future}


\bibliographystyle{IEEEtran}
\bibliography{./quantum_programming}


\end{document}

%% file: frontmatter-sv.tex
\title{Towards A High-Performance Quantum Data Center Network Architecture}
\thanks{
}
\author{\IEEEauthorblockN{Yufeng Xin}
\IEEEauthorblockA{
RENCI, University of North Carolina at Chapel Hill\\
Chapel Hill, NC, USA\\
}
\and
\IEEEauthorblockN{Liang Zhang}
\IEEEauthorblockA{
ESnet, Lawrence Berkeley National Laboratory\\
Berkeley, CA, USA
}
}

\maketitle
\thispagestyle{empty}

%% file: abstract-rev00.tex
Quantum Data Centers (QDCs) are needed to support large-scale quantum processing for both academic and commercial applications. While large-scale quantum computers are constrained by technological and financial barriers, a modular approach that clusters small quantum computers offers an alternative. This approach, however, introduces new challenges in network scalability, entanglement generation, and quantum memory management. In this paper, we propose a three-layer fat-tree network architecture for QDCs, designed to address these challenges. Our architecture features a unique leaf switch and an advanced swapping spine switch design, optimized to handle high volumes of entanglement requests as well as a queue scheduling mechanism that efficiently manages quantum memory to prevent decoherence. Through queuing-theoretical models and simulations in NetSquid, we demonstrate the proposed architecture's scalability and effectiveness in maintaining high entanglement fidelity, offering a practical path forward for modular QDC networks.

%% file: introduction-rev00.tex
With the rapid advancement of quantum computing, Quantum Data Centers (QDCs) are gaining momentum and being established globally for both academic and commercial purposes. Major technology companies like IBM and Amazon are developing QDCs for business operations, while government agencies such as the DOE and NSF are funding universities and national laboratories to set up QDCs for research. These centers are typically centered around large-scale quantum computers, often featuring hundreds of qubits. However, as quantum applications demand more qubits, building larger quantum computers has become increasingly challenging due to technological constraints and prohibitive costs.

To address this bottleneck and accelerate the deployment of QDCs, an alternative approach has recently gained traction: a modular design that connects multiple small-scale quantum computers in clusters, similar to traditional data center architecture. These smaller quantum computers, which typically have tens of qubits and are built using more mature technologies, are more cost-effective and readily available in the market. This approach offers the potential to achieve QDC performance comparable to that of large-scale quantum computers, but at a lower cost and with greater flexibility.

However, this modular QDC design also introduces new challenges, particularly in the network connecting the distributed quantum computers. In this paper, we intend to address three major challenges in QDC networks. The first challenge is scalability. Quantum networks face unique limitations from entanglement decay over distance and short qubit coherence times. Scaling Quantum Data Centers (QDCs) requires a network topology that maintains high-fidelity connections with minimal error correction and optimized routing to manage coherence constraints effectively. 

The second challenge involves entanglement generation and distribution. Quantum repeaters and switches help extend the reach of entangled quantum computers while preserving required fidelity, yet the swapping processes in these components are complex and probabilistic, affecting end-to-end entanglement fidelity and overall network throughput. 

Finally, efficient management of entangled qubit ({\bf ebit}) pairs  in quantum switch memories is critical. Given that ebits are scarce resources, scheduling them with priority and efficiency is essential to mitigate decoherence and optimize their utilization across the network.

In this paper, we present a three-layer quantum tree network architecture for Quantum Data Centers (QDC), offering a scalable solution that overcomes the limitations of traditional quantum network topologies. Alongside this tree structure, we introduce two key innovations to enhance entanglement efficiency: (1) We design an advanced leaf switch at the tree network’s bottom layer, optimized to manage high volumes of entanglement requests. This switch uses adaptive routing protocols that dynamically allocate entanglement paths, effectively preventing bottlenecks at the leaf nodes. (2) To maximize quantum memory utilization, we introduce a queue scheduling mechanism that prioritizes quantum memory requests based on factors such as coherence time, network congestion, and the importance of specific entanglement tasks. This intelligent scheduling minimizes decoherence risk and improves overall network throughput.

The remainder of this paper is organized as follows. We propose a QDC network architecture based on a spine-leaf topology in Section~\ref{sec:problem}. In Section~\ref{sec:solution}, we present our novel designs for quantum spine and quantum leaf switches within the proposed QDC architecture. We also introduce two queuing-theoretical models to evaluate the entanglement efficiency and throughput performance of these switches. In Section~\ref{sec:evaluation}, we present simulation results based on the queuing models, highlighting the key characteristics of the QDC design. Additionally, we provide a comprehensive performance evaluation of QDC networks at realistic scales using the NetSquid quantum simulator. The paper is concluded in Section~\ref{sec:future}.

%% file: related-work.tex
{\bf Quantum Entanglement Distribution networks.} A quantum network consists of three essential hardware elements: the quantum communication channel, the end quantum nodes, and the network devices. The quantum network device is needed to extend the reach of the quantum communication channels that are implemented in lossy optical channels~\cite{quantum_internet_vision_Wehner18}. Due to the inherent decoherence and non-cloning nature of qubits, development of the quantum network devices, 
such as the repeater or switch, is still in the infant stage. Earlier network layer studies focused on regular network topologies such as linear repeater chains~\cite{briegel1998quantumrepeaterscommunication}, star topologies~\cite{Epping_2017}, and lattice~\cite{Zwerger_2012}. Most recent studies expand to more general topologies 
that include tree~\cite{choi2022congestion} and mesh~\cite{caleffi2024distributed}. The main interests are the entanglement distribution and routing algorithms under realistic end-to-end 
fidelity constraints. These studies targeted general quantum architecture with two common high-level abstractions: the quantum link is responsible for entanglement generation, and the network 
devices for distribution are homogeneous repeater-like devices conducting entanglement swapping.

{\bf Performance modeling and simulation.} One line of recent research focuses on obtaining the theoretical capacity upper bound using conventional stochastic process methods of single quantum switch~\cite{switch_model_tqe_21, panigrahy2023capacity}. The capacity is defined as the maximum possible number of bipartite entanglement served by the switch per time unit. Another metric of interest, the expected number of qubits in memory at the switch $E[Q]$, was also studied. Extensive research has also been conducted in designing entanglement routing and purification schemes. Network flow optimization methods~\cite{vardoyan2024bipartite} or event-based simulation are common choices for performance studies~\cite{li2023dynamic}.

To the best of our knowledge, this paper is the first to propose a scalable quantum data center network architecture for distributed quantum computing.

%% file: problem.tex
The design of a quantum data center (QDC) network architecture should be driven by three key performance metrics: network resource constraints, quality of the entanglement when used, and the ebit utilization.

Existing quantum network architecture typically assumes an abstract quantum channel model that possesses two complex functions: heralded entanglement generation and ebit distribution. Quantum switches, or repeaters, perform entanglement swapping to extend the reach of end-to-end entanglement. However, the performance of these networks is fundamentally limited by two key quantum physical constraints: the low swapping success probability (approximately $50\%$) at swapping stations, and (2) the exponential fidelity degradation of entanglement over both distance and time. The former makes it difficult to establish entanglement paths over multiple hops, while the latter implies that entanglement connections degrade quickly, often becoming unusable in short time frames.

In addition to these inherent limitations, the current quantum network model overlooks other practical concerns, such as the high cost of heralding stations, the protocol complexity of distributing entanglement, and the inflexibility of centralized entanglement generation schemes. More critically, the abstraction of quantum links in these architectures eliminates the possibility of ebit storage, forcing unused ebits to be discarded. This results in an unacceptable waste of the most valuable resource in a quantum network. Consequently, the capacity and performance of existing quantum network architectures are insufficient to support meaningful quantum applications.  

\begin{figure}[h]
	\centering
	\includegraphics[width=0.48\textwidth]{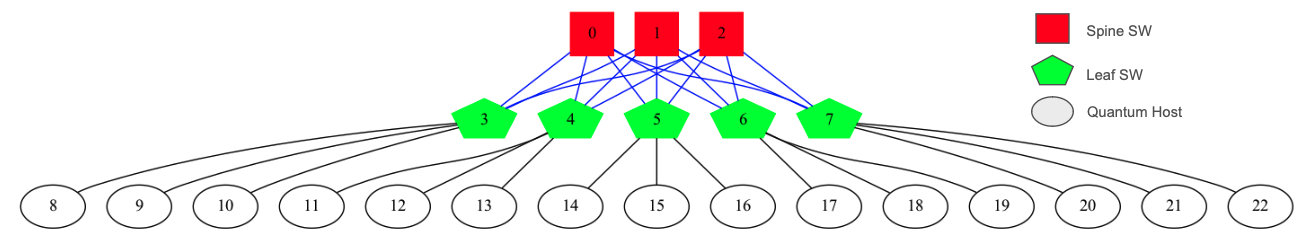}
	\caption{Quantum Data Center Network Architecture}
	\label{fig:qdc}
\end{figure}

Our first design innovation addresses these limitations by moving the ebit generation function from the quantum links back to the quantum switches. This leads to two critical performance improvements: (1) continuous generation of ebits and (2) the storage of ebits for future use. Recent work has advocated for deploying central nodes to facilitate entanglement distribution and heralding~\cite{panigrahy2023capacity,avis2023analysis}. These central nodes function primarily as centralized entanglement heralding stations, with specially designed protocols to support both bipartite and multipartite entanglement distribution. In contrast, our switch design further centralizes both ebit generation and distribution within the network, leading to better performance and cost efficiency. Additionally, recent work demonstrating a prototype quantum repeater node capable of single-photon emission for entanglement establishment across two fibers reinforces the physical viability of this design~\cite{krutyanskiy2023telecom}.   

Our second design decision is to separate the ebit swapping function from the generation function entirely, eliminating the need for ebit storage at the hosts. This separation leads to a more scalable network architecture and significantly improves network throughput. Classical data center networks have converged on a spine-leaf fat-tree topology to provide ultra-high capacity and speed. Our proposed QDC architecture naturally aligns with this topological structure. As illustrated in Fig.~\ref{fig:qdc}, the QDC architecture consists of three layers of nodes: spine switches, leaf switches, and hosts.

The hosts (represented by white boxes) are QPUs organized into clusters. Each cluster is interconnected by a leaf switch (green boxes), which facilitates entanglement connections between the hosts. A small number of spine switches (red boxes) reside at the top layer to interconnect the leaf switches and reliably enable entanglement connections between clusters. Although not shown in the diagram, a separate control network using classical channels is assumed to connect all the nodes for management and control purposes, with only quantum links displayed in the figure. The leaf and spine switches are specially designed to handle quantum operations, particularly entanglement generation and distribution.

The quantum leaf switch, the most critical and complex component of the proposed QDC architecture, is equipped with local heralded entanglement generation and storage, as well as on-demand entanglement distribution and routing capabilities. This allows for the establishment of end-to-end entanglement connections between any pair of end hosts in the network. The spine switch, in contrast, is equipped only with the entanglement swapping capability, which is activated solely for cross-cluster entanglement demands. The hosts are dedicated to executing quantum computing tasks and dynamically requesting and utilizing the required ebits as needed.

%% file: solution.tex
\subsection{Quantum Leaf Switch (Q-Leaf)}
\begin{figure}[h]
	\centering
	\includegraphics[width=0.48\textwidth]{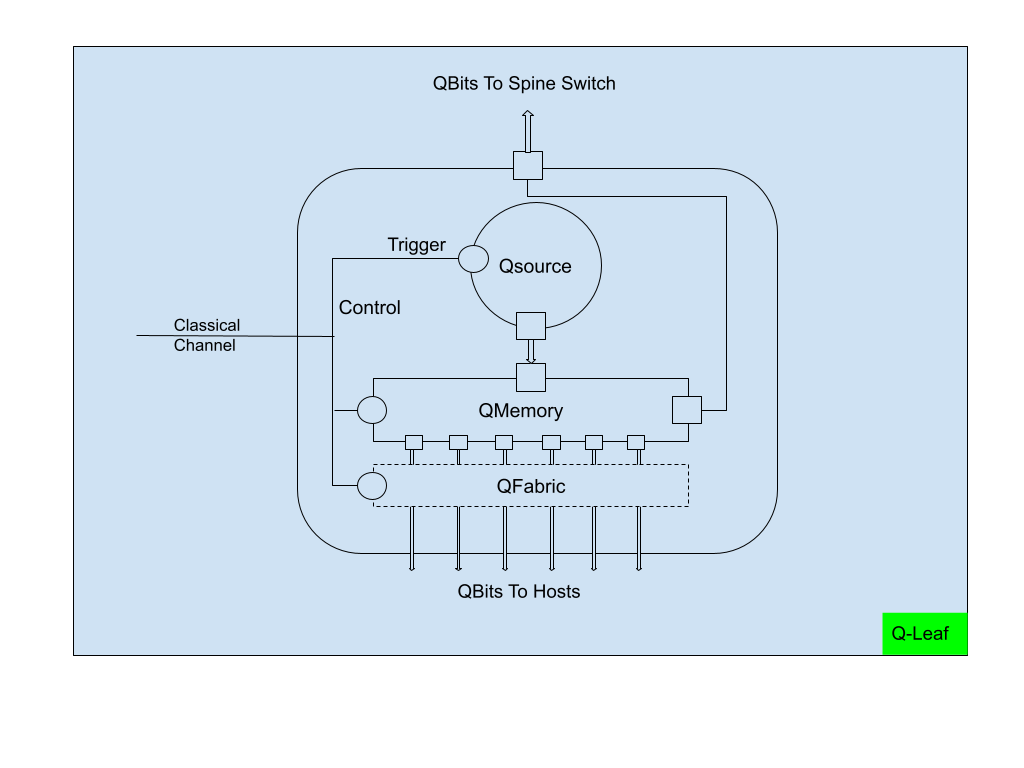}
    \vspace{-0.3in}
	\caption{Quantum Leaf Switch Architecture}
	\label{fig:q_leaf}
\end{figure}

Fig.~\ref{fig:q_leaf} shows the functional structure of the proposed leaf switch that consists of a quantum data plane and a classical control plane.

The Q-leaf switch features a centralized ebit source, a qubit memory, and a qubit fabric with ports to the hosts. This new architecture greatly simplifies 
the operational structure of the switch and significantly boost the switch performance in terms of throughput and buffer size while offering more control.
Each Q-leaf switch serves $C$ hosts in a star topology. 

\begin{figure}[h]
	\centering
	\includegraphics[width=0.48\textwidth]{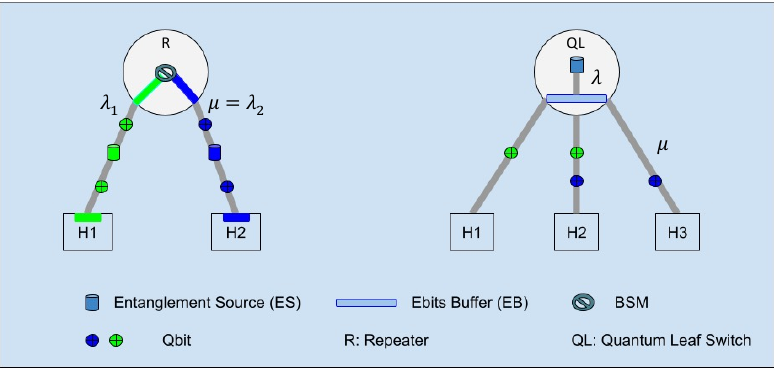}
	\caption{Quantum Switch Queue Model}
	\label{fig:q_leaf_queue}
\end{figure}

All existing quantum switch proposals assume each of the $C$ links between the hosts and the switch needs to continuously generate ebits. For each ebit, one qubit is buffered at the switch and the other is buffered at the host.  This architecture can be better represented with a queueing model. As shown in the left figure in Fig.~\ref{fig:q_leaf_queue}, one link plays the role of arrival process with rate $\lambda_1$ and the other link plays the role of the consumer with rate $\lambda_2$. In reality, the request between any pair of hosts would arrive as a stochastic process instead of the always-on assumption in the capacity upper bound model. A request has to wait until two ebits, one from each side of the repeater, become available.

As shown in the right figure in Fig.~\ref{fig:q_leaf_queue}, our proposed leaf switch can be abstracted as a capacity $M/M/C/K$ queue with reneging. When considering the dephasing of the qubits in the buffer, an ebit in the buffer will expire after its fidelity drops below a certain threshold and will leave the queue, \ie reneging. The buffer size is $K$, the number of ebits that can be stored in the memory. $C$ represents the number of servers in the queue. Contrary to the common network queue model, in this quantum switch queue model, the customer arrival process is the ebit generation process while the connection requests assume the role of servers in the queue. 

According to~\cite{chandra2022scheduling}, the fidelity loss of a Bell pair  in a quantum memory due to dephasing is given by

\begin{equation}
F(t) = \frac{1+e^{-2\Gamma t}}{2}
\end{equation}
where $\Gamma$ is the dephasing rate of the memory. 

With a targeted fidelity $F$ that is required by targeted quantum applications, the tolerable dephasing time $T$ can be derived as
 \begin{equation}
T = - \frac{\ln{(2F-1)}}{2\Gamma}
\label{renege:time}
\end{equation}
Any {\it ebit} stored in the memory longer than time $T$ will be dropped, \ie, $T$ is the reneging time in the queueing model.

Instead of the upper bound capacity, a more meaningful metric is the throughput of the switch. Using the definition from queue theory, we define the throughput as the rate at which customers, ebits, depart from the switch, as

\begin{equation}
\beta = (1-P(0))\mu
\end{equation}
$P(0)$ is the probability that the queue is empty.  

\subsection{Quantum Spine Switch}

\begin{figure}[h]
	\centering
	\includegraphics[width=0.48\textwidth]{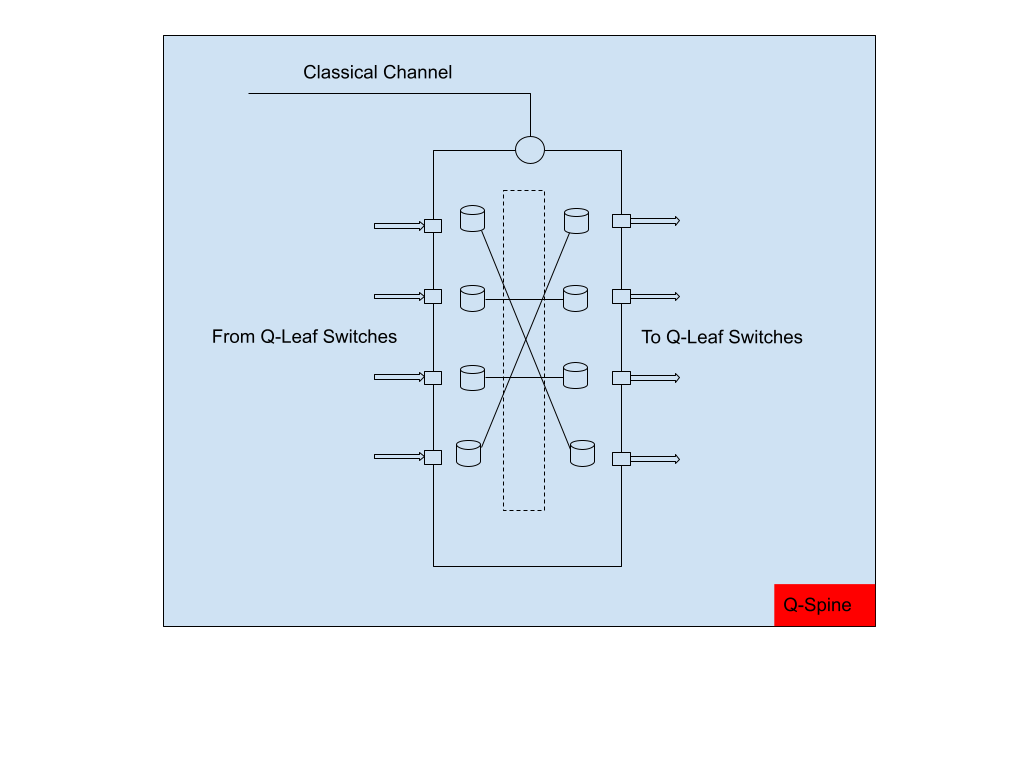}
    \vspace{-0.2in}
	\caption{Quantum Spine Switch Architecture}
	\label{fig:q_spine}
\end{figure}

Fig.~\ref{fig:q_spine} depicts the structure of the proposed spine switch. The spine switch is designed as an {\it ebit} swapping station to facilitate the entanglement connection between two hosts under different leaf switches. Upon such a connection request, the two corresponding leaf switches will each send a {\it qubit} out of an {\it ebit} to a spine switch for a swapping operation while sending the other {\it qubit} to the hosts respectively. 

\begin{figure}[h]
	\centering
	\includegraphics[width=0.48\textwidth]{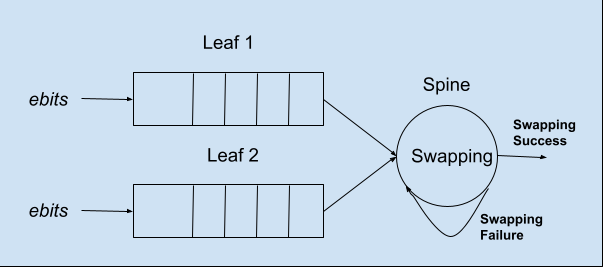}
	\caption{Spine Switch Assembly Queue System Model}
	\label{fig:q_spine_queue}
\end{figure}

For performance analysis, we model the spine-leaf network as a product assembly line system, where the swapping operation is modeled as an assemble service that 
needs two {\it ebits}, the parts, available. The main {\it ebit} swapping operation involves a BSM (Bell State Measurement) that is subject to probabilistic failure, which is represented by the assembly failure probability in the model. A swapping failure automatically discards the current {\it ebit} pair and tries the next pair of available {\it ebits}. The queue model for one spine switch with two leaf switches is depicted in Fig.~\ref{fig:q_spine_queue}.

%% file: evaluation-rev00.tex
We evaluate the performance of the proposed QDC architecture with two separate simulation studies. A queueing theoretical model-based simulation is conducted using the Simpy event simulator for a single leaf switch cluster and a network with a single spine switch. A NetSquid-based simulation for Quantum Data Center (QDC) networks was developed to evaluate performance under realistic quantum constraints and device conditions~\cite{netsquid}.

\subsection{Queueing Model Simulation}
We first study the performance of a QDC leaf switch with three hosts using the M/M/c/K with reneging model. The hosts generate pairwise service requests with a given service rate. In the simulation, we assume the fidelity threshold to be $0.7$, which decides when an ebit will renege according to Equation~\ref{renege:time}. The ebit generation rate is set to 30 per time unit. 

\begin{figure}[h]
	\centering
	\includegraphics[width=0.48\textwidth]{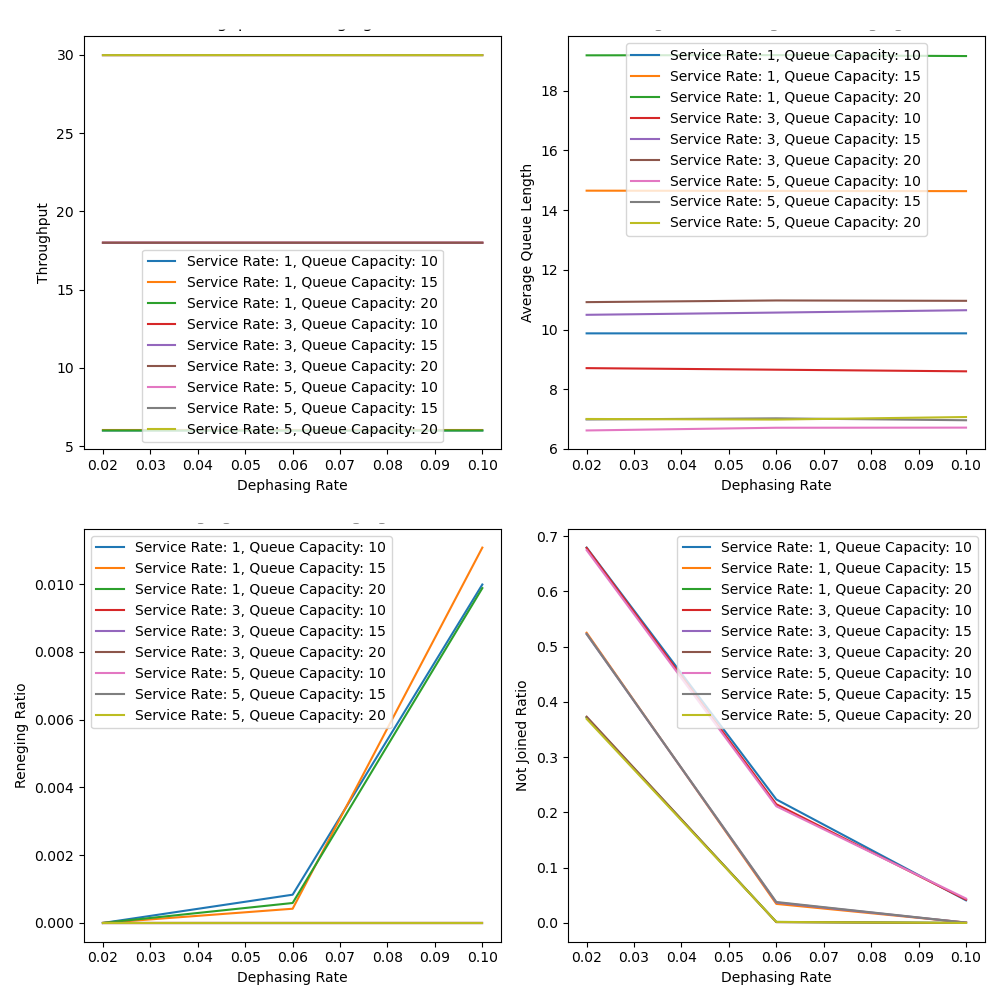}
	\caption{Q-Leaf Switch Performance (ebit generation rate = 30)}
	\label{fig:q_leaf_30}
\end{figure}

In Figure~\ref{fig:q_leaf_30}, we depict four performance metrics: throughput, average queue length, reneging ratio, and not-joined ratio against the dephasing rate ranging between 0.02 and 0.1. In each figure, we show six sets of results with queue capacity ranging $\{10, 15, 20\}$, service rate per host pair ranging in $\{1.0, 1.5, 2.0\}$ for generation rate 30. We recall that the ebit generation rate, queue capacity, and reneging are largely decided by quantum engineering. Small quantum buffers and low reneging are highly preferable for both engineering and cost considerations.

The throughput result shows that the leaf switch can satisfy high volumes of requests from the hosts as long as the ebit generation can keep up, independent of the dephasing performance of the qubit memory. The average queue length performance figures show that it is possible to only use quantum memory of small size and long queueing time can still be avoided. 
Unsurprisingly, the next figure shows that reneging rate is greatly affected by the dephasing performance and increases significantly after the dephasing rate is beyond $0.06$, especially when the service rate is low. On the other hand, higher dephasing leads to a lower not-joining rate as the memory is refreshed quickly due to sooner reneging time. These results signify the need to quantify the tradeoff between the costly {\it ebit} generation and buffering capacity constraints and performance under certain demand profiles.

We next present the simulation results on a basic QDC network consisting of 1 spine, 2 leaf switches, and 3 hosts per leaf switch with the assembly line queue model. We are specifically interested in the successful assembly rate under the impacts of the spine switch parameters, \ie successful assembly probability and the leaf switch parameters, dephasing rate ($d_r$), buffer capacity, {\it ebit} generation rate (arrival rate), and the demand rate from the hosts ($a_r$). In the simulation, $40\%$ of the demands are inter-cluster demands that need to go through the swapping processing in the spine switch and the rest $60\%$ demands are intra-cluster demands that are between two hosts under the same leaf switch. 

\begin{figure}[h]
	\centering
	\includegraphics[width=0.48\textwidth]{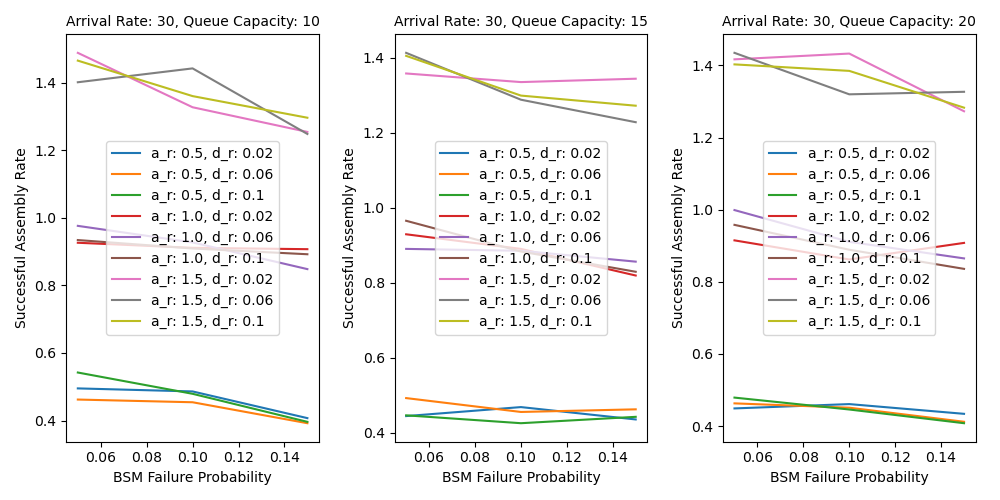}
	\caption{Q-Spine Switch Performance (ebit generation rate = 30)}
	\label{fig:q_spine_30}
\end{figure}

We present the results in Fig. ~\ref{fig:q_spine_30} with the ebit generation rate set to 30 per time unit in the two leaf switches. 
Each figure consists of three subfigures generated with different buffer sizes at the leaf switch of 10, 15, and 20 respectively. Each subfigure aggregates results from leaf switch parameter combinations of service rate ranging in 0.5, 1.0, and 1.5 and dephasing rate in 0.02, 0.06, and 0.1. As expected, a higher BSM failure rate negatively affects the swapping success rate, especially for the cases of higher connection request rates ($a_r$). The results also indicate that higher queue capacity helps some when $a_r$ and the {\it ebit} generation rate are higher. The impacts of these parameters indicate complex tradeoff relationships to the system performance. Nevertheless, this simulation model provides convenient guidance to the switch design under realistic quantum engineering constraints.   

\begin{figure}[h]
	\centering
	\includegraphics[width=0.48\textwidth]{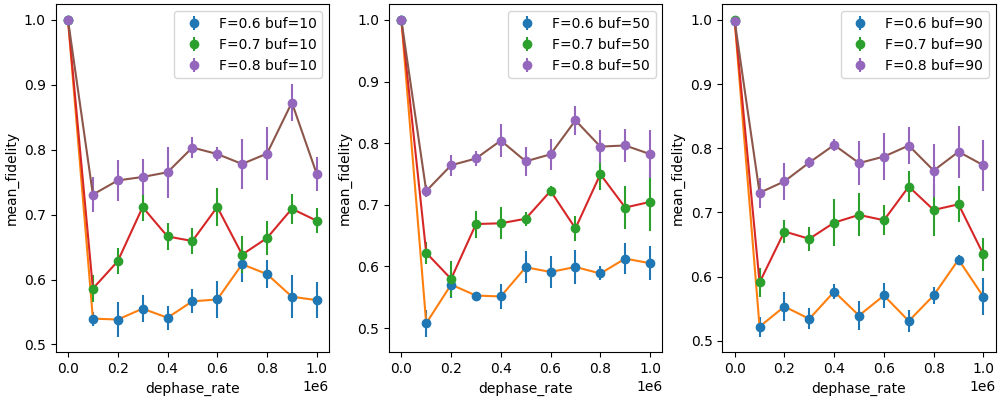}
	\caption{Fidelity vs Dephase Rate}
	\label{fig:netsquid-dephase-fidelity}
\end{figure}

\begin{figure}[h]
	\centering
	\includegraphics[width=0.48\textwidth]{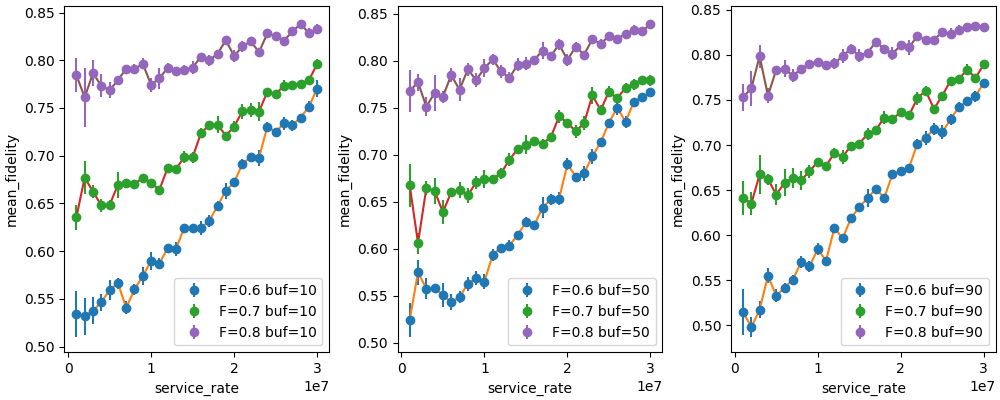}
	\caption{Fidelity vs Service Rate}
	\label{fig:netsquid-servicerate-fidelity}
\end{figure}

\begin{figure}[h]
	\centering
	\includegraphics[width=0.48\textwidth]{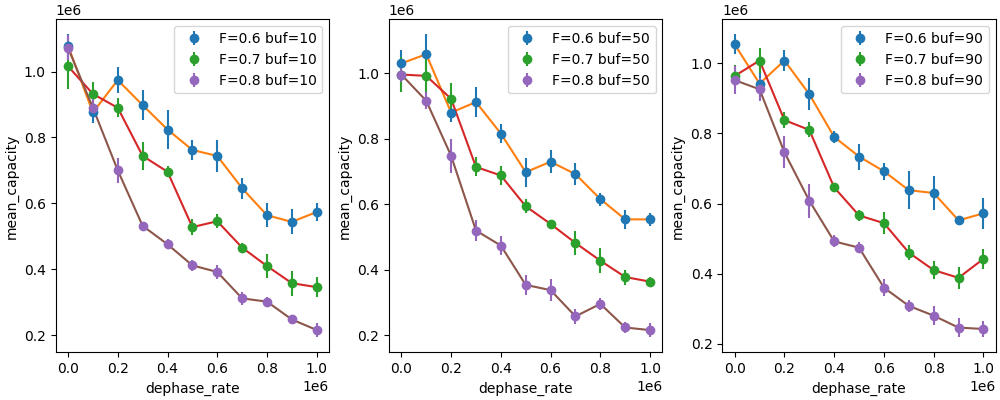}
	\caption{Capacity vs Dephase Rate}
	\label{fig:netsquid-dephase-capacity}
\end{figure}

\begin{figure}[h]
	\centering
	\includegraphics[width=0.48\textwidth]{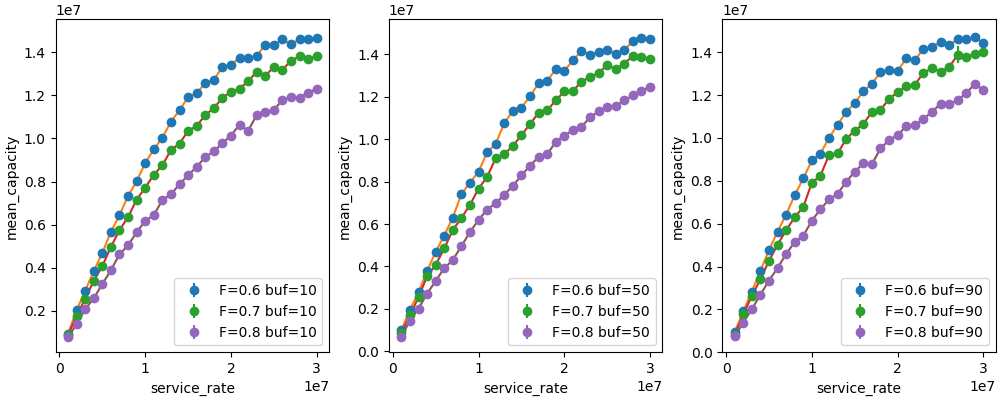}
	\caption{Capacity vs Service Rate (Dephase Rate = 0.2)}
	\label{fig:netsquid-servicerate-capacity-2}
\end{figure}

\subsection{Netsquid Simulation}

We implemented the QDC network using NetSquid and evaluated critical metrics, including fidelity, capacity, and scalability. The QDC network operates as follows: entangled qubit pairs are generated locally at the leaf switches, where they are stored in quantum memories managed by a priority queue. When a service request for entanglement is received between two end hosts, one entangled pair is selected from the memory and distributed to the remote hosts immediately. Due to decoherence, the qubit pair in memory will be discarded once their states decay below the required fidelity threshold \( F \) after a certain period defined in formula (2). Link noise is disregarded, given the short distances between switches and hosts within data centers. Upon reaching the remote hosts, the fidelity of the qubit pair is calculated against its initial state and included in the statistical analysis. 

\textbf{Fidelity in Relation to Dephasing and Buffer Size} We examined the relationship between the fidelity of received qubits and multiple influencing factors, as depicted in Fig.~\ref{fig:netsquid-dephase-fidelity}. In our evaluation, we assume that decoherence arises solely from dephasing. The x-axis in each figure represents the dephasing rate (Hz), with entangled qubit pair being generated at a rate of 1 MHz on each leaf switch. Different subplot figures denote varying buffer sizes, ranging from 10 to 90. Our findings indicate that fidelity changes gradually with varying dephasing rates, except when the dephasing rate is zero. Additionally, increasing the buffer size does not significantly impact the fidelity of the received qubits. Each line in Fig.~\ref{fig:netsquid-dephase-fidelity} corresponds to different values of \( F \), specifically 0.6, 0.7, and 0.8, illustrating that \( F \) critically determines the fidelity range observed in each scenario.

These experiments demonstrate that a relatively small amount of quantum memory can achieve fidelity levels comparable to those achieved with larger memory sizes. The primary determinant of fidelity appears to be the quantum queuing system, especially the \( F \) value, which regulates the duration that qubits remain in memory before being discarded.

\textbf{Fidelity in Relation to Service Rate}
Fig.~\ref{fig:netsquid-servicerate-fidelity} illustrates the fidelity of received qubits relative to their initial states. The data shows a clear trend: the fidelity of received qubits increases with higher service rates. This is because higher service rates result in more recently generated qubits being transmitted, reducing the likelihood of decoherence and other errors that can degrade quantum states.

\textbf{System Capacity in Relation to Dephasing and Buffer Size} Fig.~\ref{fig:netsquid-dephase-capacity} illustrate the capacity of the Quantum Data Center (QDC) in relation to the dephasing rate of the quantum memory. In each figure, different colors represent various \( F \) values at the leaf switches. As the dephasing rate increases, the overall capacity of the QDC decreases. This trend is consistent across all figures. The capacity diminishes more rapidly with higher dephasing rates because faster decoherence causes the fidelity of the qubits to degrade below acceptable \( F \) more quickly, resulting in fewer qubits being usable for transmission. Higher \( F \) values exacerbate the reduction in capacity. The figures show that when higher fidelity is required, the system discards more qubits that fail to meet these stringent criteria. This demonstrates that stricter fidelity requirements lead to a significant drop in the number of qubits that can be effectively used, thereby lowering the system's capacity. The data indicate that increasing buffer sizes from small to large does not significantly impact the system's capacity. The capacity remains relatively stable irrespective of the buffer size. This suggests that using smaller queue memory sizes is sufficient to meet performance requirements, providing an efficient use of resources without sacrificing capacity.

The analysis of Fig.~\ref{fig:netsquid-dephase-capacity} highlights that the primary factors affecting QDC capacity are the dephasing rate and the fidelity threshold \( F \), while buffer size has a minimal impact. Minimizing dephasing and controling tolerable fidelity thresholds are critical for designing efficient quantum data centers that can maintain high capacity and performance under varying operational conditions.

\textbf{System Capacity in relation to Service Rates} In this study, we examine the system capacity as a function of service rates under varying conditions of decoherence. The experimental setup involves 15 leaf switches, each generating entangled qubits at a rate of 1 MHz. The service requests are randomly generated and evenly distributed across all leaf switches. We investigate the performance across different decoherence rates: 0, 0.2, and 0.5 times the entanglement rate, with service rates ranging from 1 to 30 times the entanglement rate. We observe that across all three settings, the system capacity plateaus beyond the 15 MHz threshold. This occurs because the number of available qubits becomes inadequate to meet the increasing service demand. In other words, the system hits its maximum capacity, unable to handle additional requests due to a shortage of entangled qubits. Fig.~\ref{fig:netsquid-servicerate-capacity-2} illustrates the setting with a decoherence rate of 0.2. The system's capacity still shows an initial linear increase with the service rate, but the slope is less steep compared to zero decoherence rate. The maximum capacity is achieved at a lower service rate, reflecting the loss of qubits due to decoherence. The impact of decoherence begins to manifest more prominently as the service rate approaches the system's upper capacity limit.

The experimental results clearly show that system capacity is heavily influenced by both service rates and decoherence rates. When decoherence is negligible, the system can scale linearly with service demand up to a certain threshold. However, as decoherence rates increase, the system's ability to handle higher service rates diminishes due to the loss of qubits. This finding underscores the importance of minimizing decoherence in quantum communication systems to maintain high system capacity and efficient scalability.

%% file: future.tex
By leveraging a scalable spine-leaf topology, our proposed quantum data center network can efficiently manage and distribute entangled qubits among a large number of quantum computers. The integration of semi-centralized {\it ebits} generation, locally buffered entanglement links, and on-demand entanglement distribution further optimizes performance, making the architecture well-suited for supporting advanced quantum applications like distributed quantum computing. 

We will further refine the physical designs of the quantum switches to ensure the QDC network can efficiently scale. We will also develop advanced network control protocols to support emerging quantum computing applications.